\documentclass[3p,times,twocolumn]{elsarticle}

\usepackage{ecrc}

\volume{00}

\firstpage{1}

\journalname{Nuclear and Particle Physics Proceedings}

\usepackage{amsmath}

\newcommand{ \be }{\begin{equation}}      
\newcommand{ \ee }{\end{equatiton}}      
\newcommand{ \bea }{\begin{eqnarray}}      
\newcommand{ \eea }{\end{eqnarray}}

\usepackage{color}

\newcommand{ \sNN }{\sqrt{s_{\text{\textrm{NN}}}} }           

\newcommand{\pT}{$p_{T}$ }
\newcommand{\pp}{p+p }
\newcommand{\dAu}{d+Au }

\newcommand{\AuAu}{Au+Au }

\newcommand{\raa}{$R_{AA}$ }
\newcommand{\ncoll}{$N_{\text{\textrm{coll}}}$ }

\runauth{}

\jid{nppp}

\jnltitlelogo{Nuclear and Particle Physics Proceedings}

\usepackage{amssymb}

\usepackage[figuresright]{rotating}

\begin{document}

\begin{frontmatter}



\dochead{}

\title{Measurement of $D^0$ Meson Production and Azimuthal Anisotropy in Au+Au Collisions at $\sNN$ = 200 GeV}


\author{Guannan Xie (for the STAR\fnref{col1}  Collaboration)}
\fntext[col1] {A list of members of the STAR Collaboration and acknowledgments can be found at the end of this issue.}

\address{
University of Science and Technology of China, Hefei, 230026, China

Lawrence Berkeley National Laboratory, Berkeley, CA 94720, USA
}

\begin{abstract}
  Due to the large masses, heavy-flavor quarks are dominantly produced in initial hard scattering processes and experience the whole evolution of the medium produced in heavy-ion collisions at RHIC energies. They are also expected to thermalize slower than light-flavor quarks. Thus the measurement of heavy quark production and azimuthal anisotropy can provide important insights into the medium properties through their interactions with the medium. 

  In these proceedings, we report measurements of $D^0$ production and elliptic flow ($v_2$) via topological reconstruction using STAR's recently installed Heavy Flavor Tracker (HFT). The new measurement of the nuclear modification factor ($R_{AA}$) of $D^0$ mesons in central Au+Au collisions at $\sNN$ = 200 GeV confirms the strong suppression at high transverse momenta ($p_{T}$) reported in the previous publication with much improved precision. We also report the measurement of elliptic flow for $D^0$ mesons in a wide transverse momentum range in 0-80\% minimum-bias Au+Au collisions. The $D^0$ elliptic flow is finite for \pT $>$ 2 GeV/\text{\textit{c}} and is systematically below that of light hadrons in the same centrality interval. Furthermore, several theoretical calculations are compared to both \raa and $v_2$ measurements, and the charm quark diffusion coefficient is inferred to be between 2 and $\sim$12.
\end{abstract}


\end{frontmatter}


\section{Introduction}
The mass of the charm quark is significantly larger than those of light quarks, the QCD scale, and the temperature of the quark-gluon plasma (QGP) created at RHIC energies. The charm quark mass is mostly unaffected by the QCD medium, and the charm quarks are dominantly produced at the early stage of heavy-ion collisions through hard scattering processes at RHIC. They experience the whole evolution of the system. Therefore, charm quarks provide unique information on the properties of hot and dense strongly-coupled QGP.

The charm quark production has been systematically studied in p+p ($\overline{\text{\textrm{p}}}$) collisions at various experiments. Figure 1 shows the charm quark differential cross-section versus transverse momentum (\pT) in p+p collisions at $\sqrt{s}= 200$ GeV-7 TeV~\cite{star1, star2, cdf, alice}. Experimental data are compared with Fixed-Order Next-to-Leading-Log (FONLL) pQCD calculations shown as grey bands~\cite{fonll}. Within uncertainties, FONLL calculations describe the data over a broad range of collision energies. At RHIC energies, charm quarks are produced mostly via initial hard scatterings. This is confirmed in Figure 2 where the total charm quark cross sections in p+p, \dAu and \AuAu collisions are shown to scale with the number of binary nucleon-nucleon collisions (\ncoll)~\cite{star1, fonll, star3, star4, nlo}. 

Recent measurements at both the RHIC and the LHC show that high \pT charmed meson production is considerably suppressed in the central heavy-ion collisions, which indicates strong interactions between charm quarks and the medium. It is also found that the $D$-meson elliptic flow measured at the LHC is comparable with that of light hadrons~\cite{alicev2}.

\begin{figure}[htbp]
\hspace{+1.2cm}
\begin{minipage}[b]{0.65\linewidth}
\begin{center}
\includegraphics[width=\textwidth]{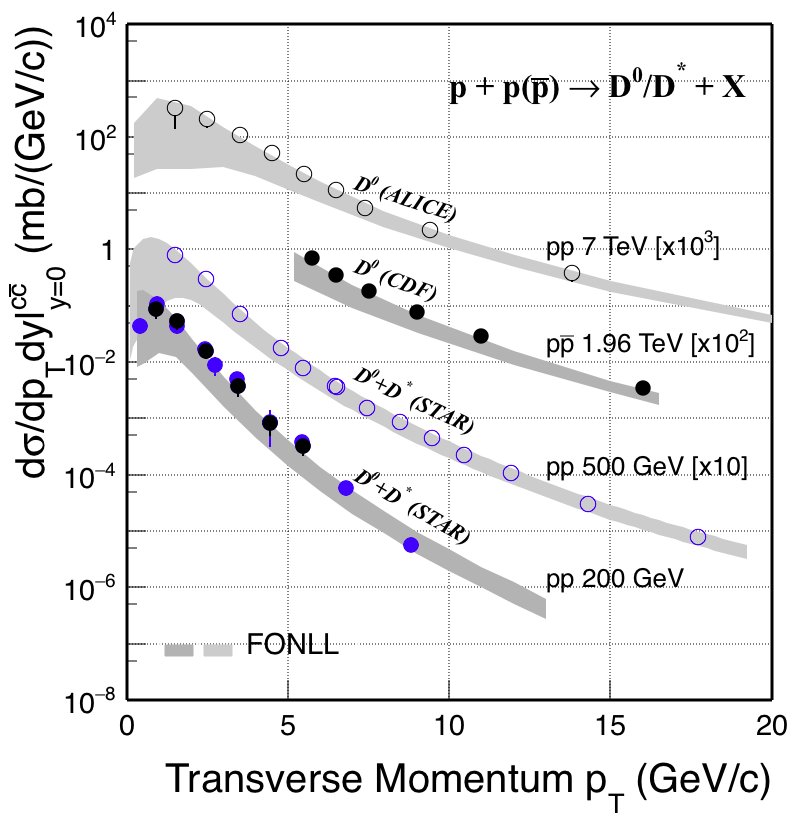}
\end{center}
\end{minipage}
\put(-75, 30) {\footnotesize \color{blue} STAR Preliminary}
\caption{ Charm quark pair production cross section vs. \pT at mid-rapidity in p+p ($\overline{\text{\textrm{p}}}$) collisions~\cite{star1, star2, cdf, alice}.} 
\label{fig:cc_pp_xSection}
\end{figure}

\begin{figure}[htbp]
\hspace{+1.2cm}
\begin{minipage}[b]{0.7\linewidth}
\begin{center}
\includegraphics[width=\textwidth]{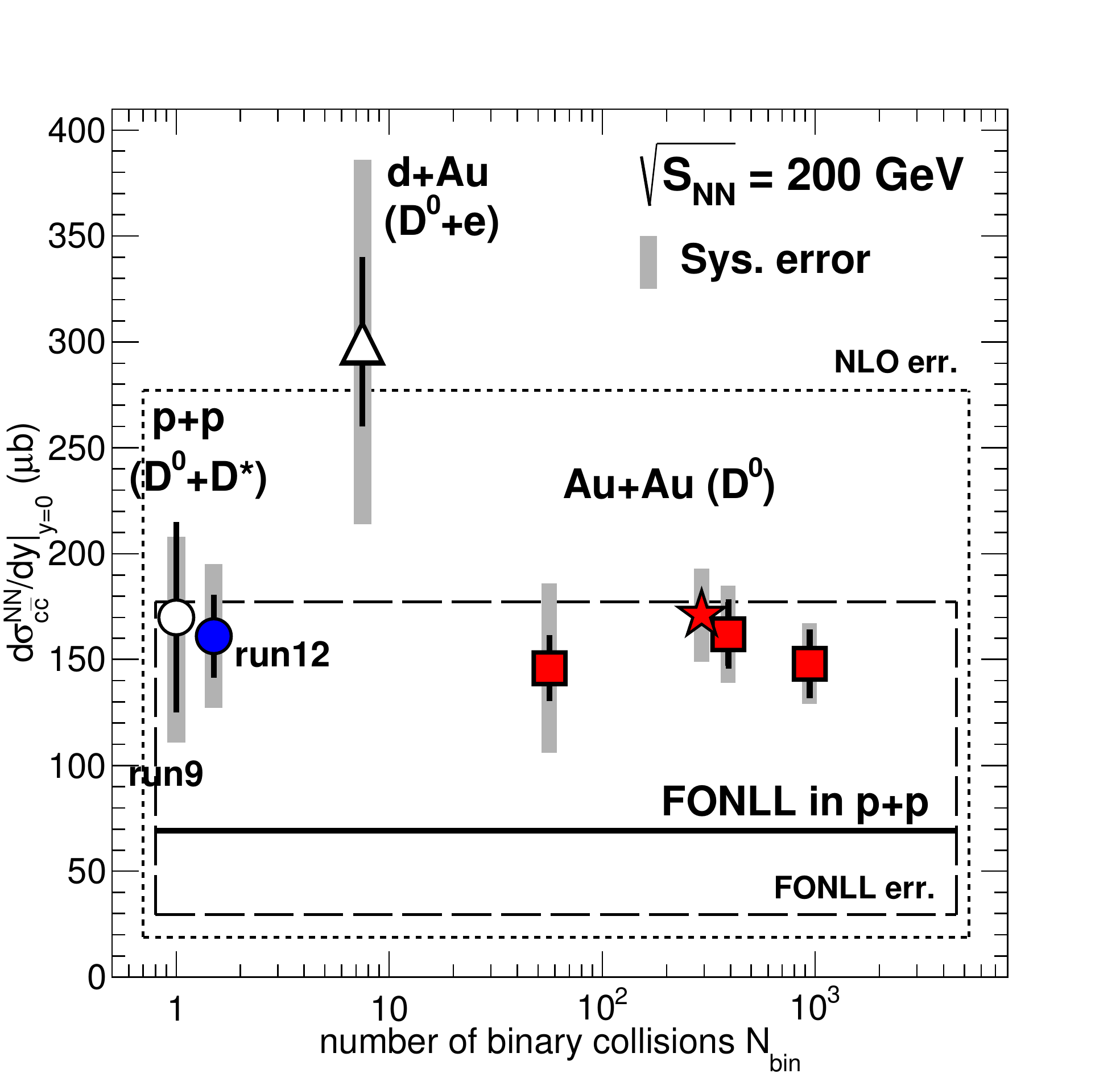}
\end{center}
\end{minipage}
\put(-115, 28) {\footnotesize \color{blue} STAR Preliminary}
\caption{ Charm quark cross sections at mid-rapidity in p+p, \dAu and \AuAu collisions measured by the STAR experiment.} 
\label{fig:cc_xx_xSection}
\end{figure}

\section{Experiment and Analysis}

The STAR experiment is a large-acceptance detector covering full azimuth and pseudorapidity of $|\eta| < 1$ at the RHIC. Data were taken by the STAR experiment using the newly installed Heavy Flavor Tracker (HFT) in the year 2014. The HFT is a high resolution silicon detector which provides a track pointing resolution of less than 50 $\mu m$ for kaons with \pT = 750 MeV/$c$.

About 780M minimum bias \AuAu events are used in this analysis. These events are selected to contain primary vertices within 6 cm to the center of the STAR detector along the beam direction for uniform HFT acceptance. $D^0$ and $\overline{D^0}$ are reconstructed through the hadronic decay channel, with a branching ratio of $\sim3.9\%$ and a lifetime of $c\tau\sim123$ $\mu m$. The kaons and pions are identified using the energy loss (dE/dx) measured by the Time Projection Chamber (TPC) and the time of flight measured by the Time-Of-Flight (TOF) detector~\cite{stardetector}. The secondary vertices are reconstructed as the middle points at the Distance of the Closest Approach (DCA) between the two daughter particles. With the HFT, the following topological cuts are applied to greatly reduce the combinational background: decay length (distance between primary and decay vertices), DCA between daughter tracks, DCA between reconstructed $D^0$ and the primary vertex, DCA between daughter tracks and the primary vertex. Topological cuts are optimized in each $D^0$ \pT bins using the Toolkit for Multivariate Data Analysis (TMVA) package to achieve the best $D^0$ signal significance.

\begin{figure}[htbp]
\hspace{+1.2cm}
\begin{minipage}[b]{0.80\linewidth}
\begin{center}
\includegraphics[width=\textwidth]{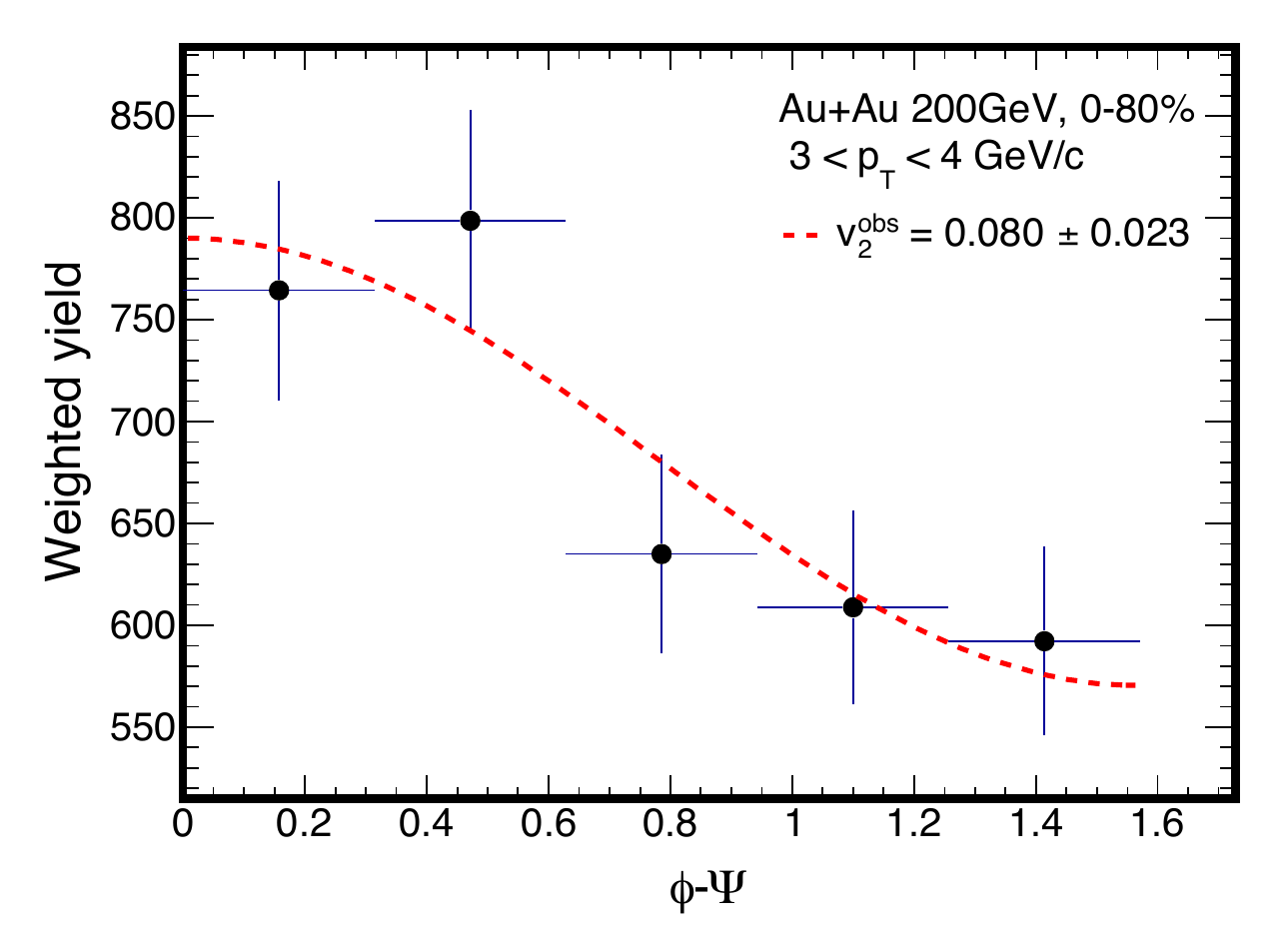}
\end{center}
\end{minipage}
\put(-135, 30) {\footnotesize \color{blue} STAR Preliminary}
\caption{ $D^0$ yield as a function of the angle relative to the event plane $(\phi - \Psi)$ for 3 $<$ \pT $<$ 4 GeV/\text{\textit{c}}, in 0-80\% \AuAu collisions.}
\label{fig:v2Methods}
\end{figure}

The event plane method is used to extract the second-order azimuthal anisotropy ($v_2$) for $D^0$. The second-order event plane ($\Psi$) is reconstructed using TPC tracks excluding decay products of $D^0$ mesons and corrected for non-uniform detector efficiency. In order to reduce the non-flow contribution, a $\eta$-gap of $|\Delta\eta| > 0.15$ between $D^0$ mesons and charged tracks used for event plane reconstruction is required. The azimuthal distribution of $D^0$ mesons with respect to the event plane $(\phi - \Psi)$ is then obtained and weighted by 1/($\epsilon$*R) for each centrality, where $\epsilon$ is the $D^0$ reconstruction efficiency and R the event plane resolution. The observed $v_2$ ($v_2^{obs}$) is obtained by fitting the distribution of $D^0$ yield versus $(\phi - \Psi)$ with a functional form of $A(1+2v_{2}^{obs}\cos(2(\phi - \Psi)))$ taking into account the finite bin width effect. Finally, the true $v_2$ is obtained by scaling $v_2^{obs}$ with $1/R$ to correct for the event plane resolution. Figure 3 shows the weighted yield as a function of $(\phi - \Psi)$ for $D^0$ candidates with 3 $<$ \pT $<$ 4 GeV/$c$. The remaining contribution of non-flow effects to the measured $v_2$ is estimated by scaling the non-flow effect in \pp collisions to Au+Au collisions~\cite{v2art}.

\begin{figure}[htbp]
\hspace{+1.2cm}
\begin{minipage}[b]{0.80\linewidth}
\begin{center}
\includegraphics[width=\textwidth]{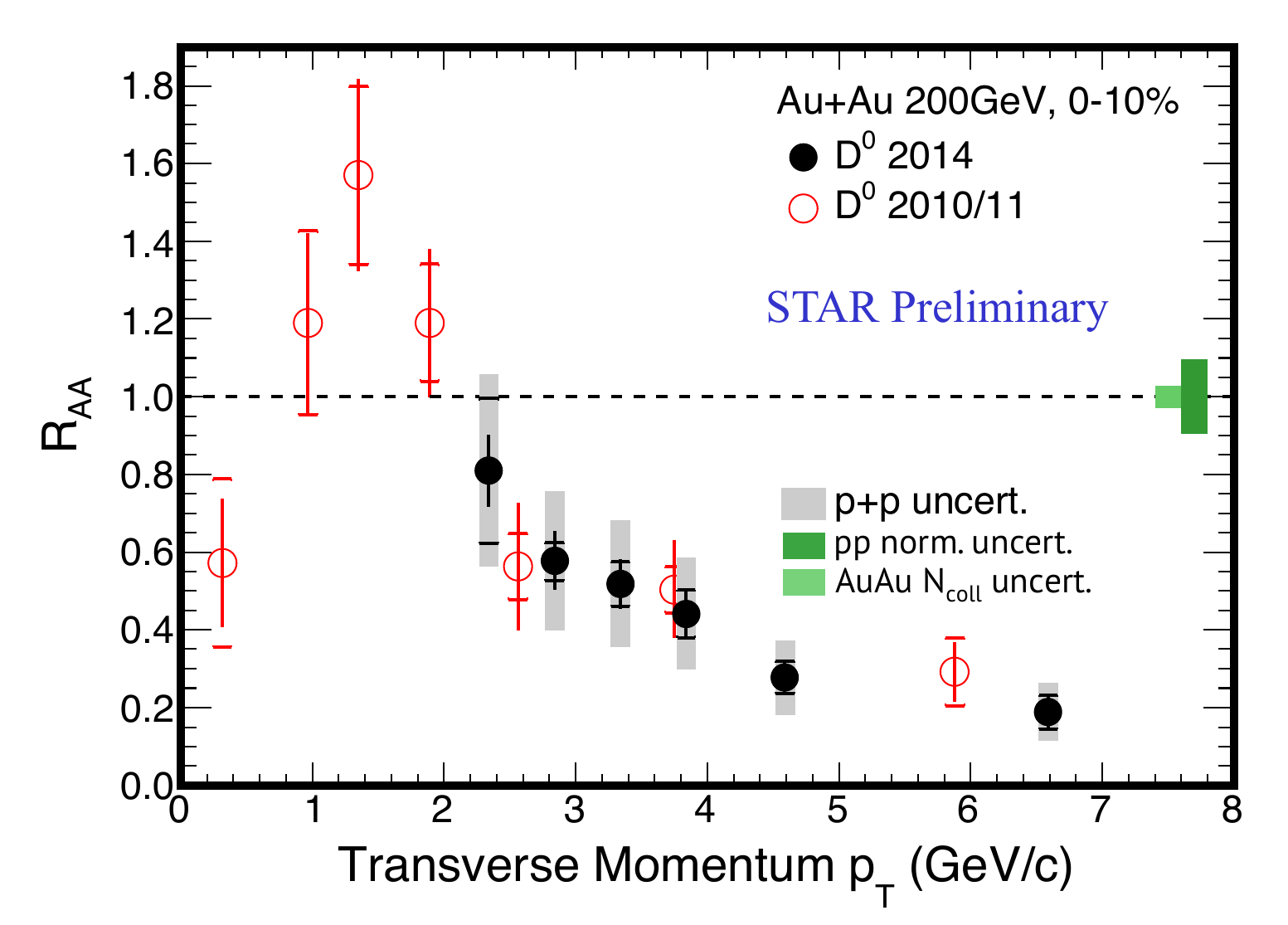}
\end{center}
\end{minipage}
\caption{ $D^0$ \raa in 0-10\% central Au+Au collisions.}
\label{fig:d0raa}
\end{figure}

\section{Physics Results and Discussion}

Figure 4 shows the \raa for the most central (0-10\%) \AuAu collisions. The new results from the HFT are consistent with the published ones above 2 GeV/\text{\textit{c}} with significantly improved precision for \AuAu measurements. The grey bands show uncertainties from the \pp baseline measured before the HFT installation~\cite{star1}. The \raa shows a strong suppression at high \pT indicating strong charm-medium interactions at this kinematic region.


\begin{figure}[htbp]
\hspace{+1.2cm}
\begin{minipage}[b]{0.80\linewidth}
\begin{center}
\includegraphics[width=\textwidth]{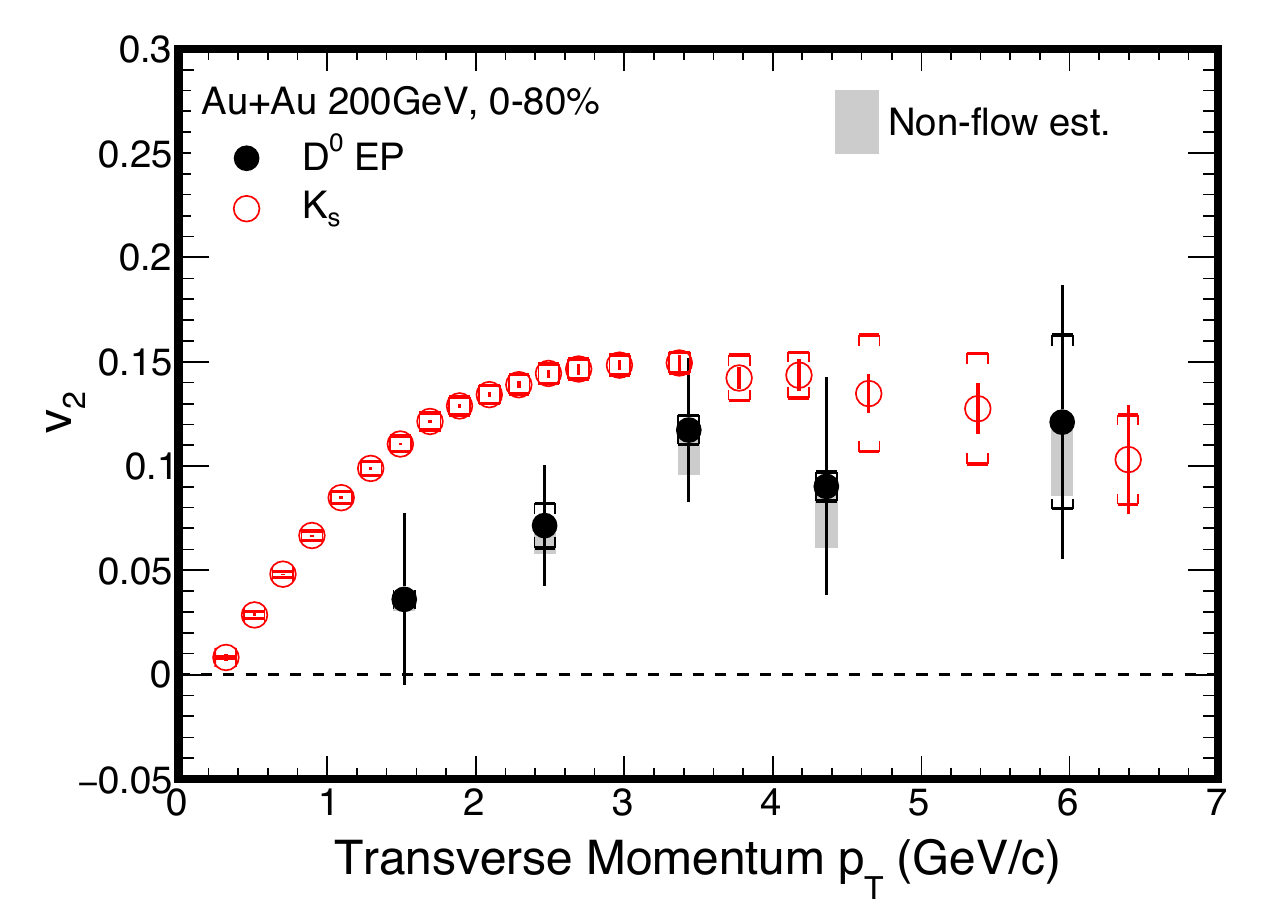}
\end{center}
\end{minipage}
\put(-135, 25) {\footnotesize \color{blue} STAR Preliminary}
\caption{ $D^0$ $v_2$ as a function of $p_T$, compared with that of $K_s$.}
\label{fig:d0v2_1}
\end{figure}

Figure 5 shows the $v_2$ of $D^0$ mesons compared to that of $K_s$. The measured $D^0$ $v_2$ is non-zero and systematically below that of $K_s$ in the range 1 $<$ \pT $<$ 6 GeV/\text{\textit{c}}. To account for the different particle masses and number of constituent quarks ($n_q$), another comparison is shown in Fig. 6, i.e. $v_2/n_q$ vs. $(m_T - m_0)/n_q$ where $m_T = \sqrt{p_T^{2}+m_0^{2}}$. After the $n_q$ scaling, even though the $D_0$ $v_2$ is still below that of light hadrons, the difference is reduced. As the comparison is done in a wide centrality range (0-80\%) in which the $D^0$ production is more biased towards central collisions than light hadrons, a fair comparison in finer centrality ranges is needed to draw firm conclusions~\cite{v2star1, v2star2, v2star3, michael}.

\begin{figure}[htbp]
\hspace{+1.2cm}
\begin{minipage}[b]{0.80\linewidth}
\begin{center}
\includegraphics[width=\textwidth]{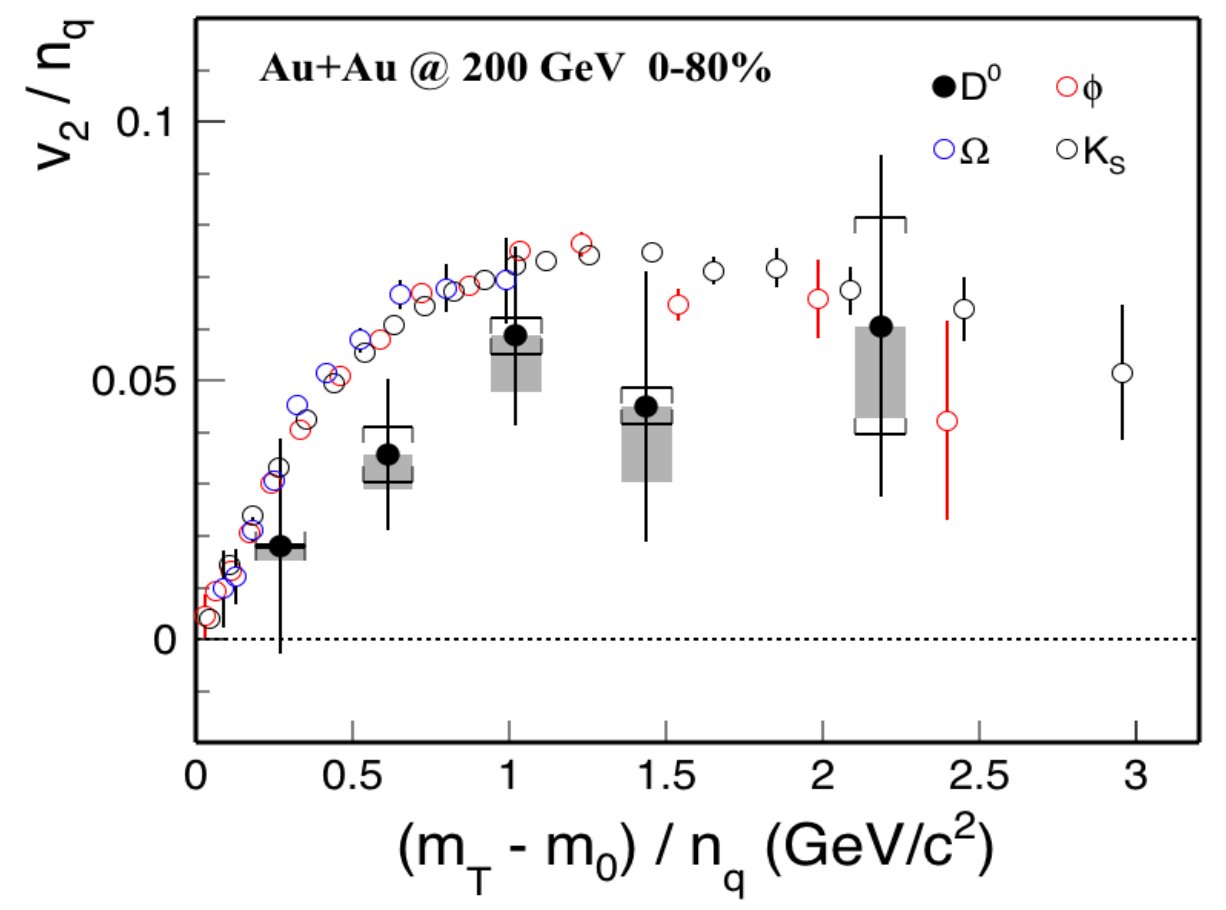}
\end{center}
\end{minipage}
\put(-135, 30) {\footnotesize \color{blue} STAR Preliminary}
\caption{ $D^0$ $v_2/n_q$ as a function of $(m_T - m_0)/n_q$ compared with that of light hadrons.}
\label{fig:d0v2_2}
\end{figure}

\begin{figure}[htbp]
\hspace{+1.2cm}
\begin{minipage}[b]{0.80\linewidth}
\begin{center}
\includegraphics[width=\textwidth]{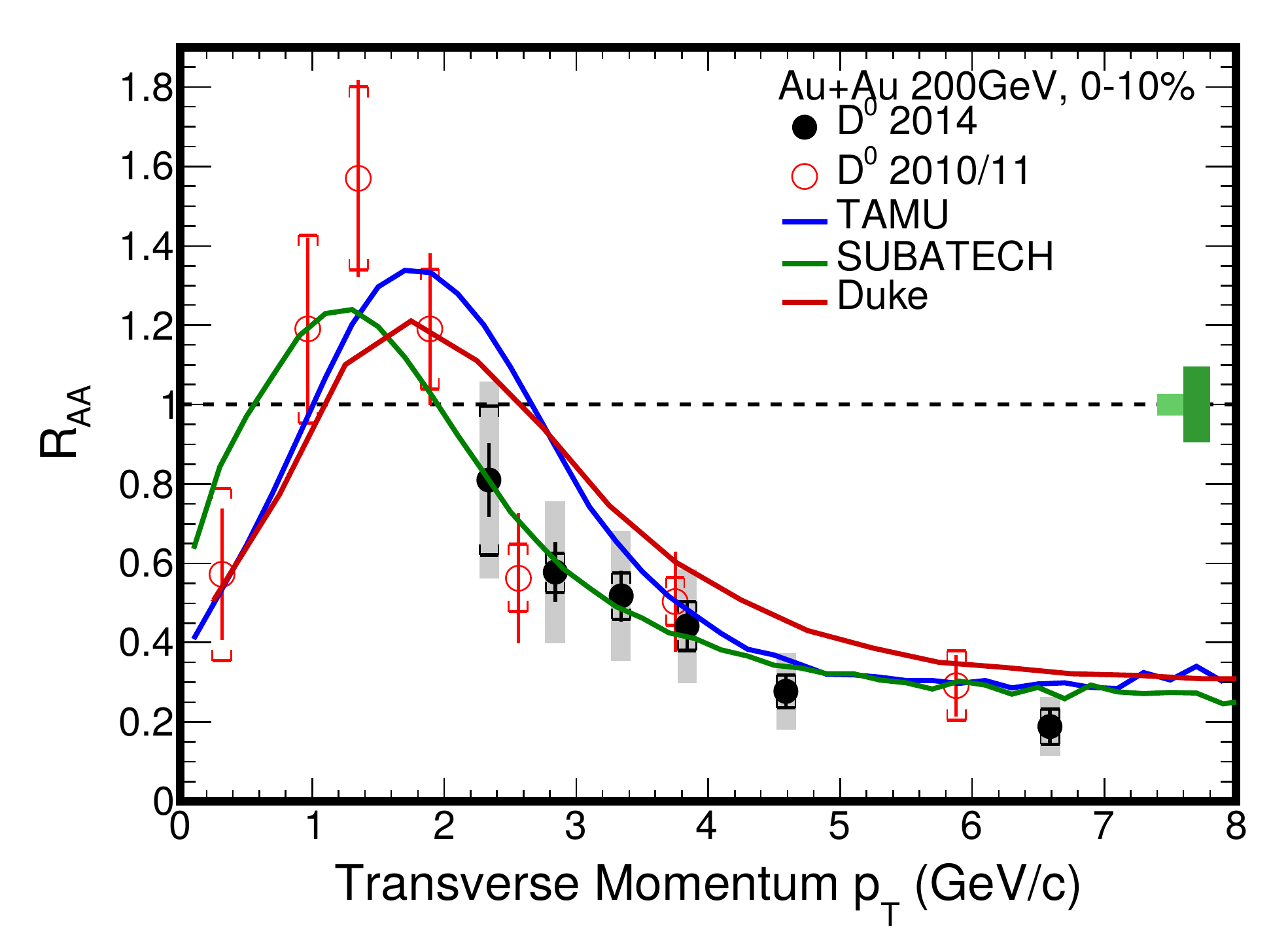}
\end{center}
\end{minipage}
\put(-135, 25) {\footnotesize \color{blue} STAR Preliminary}
\caption{ $D^0$ \raa compared to various models.}
\label{fig:d0v2_2}
\end{figure}

\begin{figure}[htbp]
\hspace{+1.2cm}
\begin{minipage}[b]{0.80\linewidth}
\begin{center}
\includegraphics[width=\textwidth]{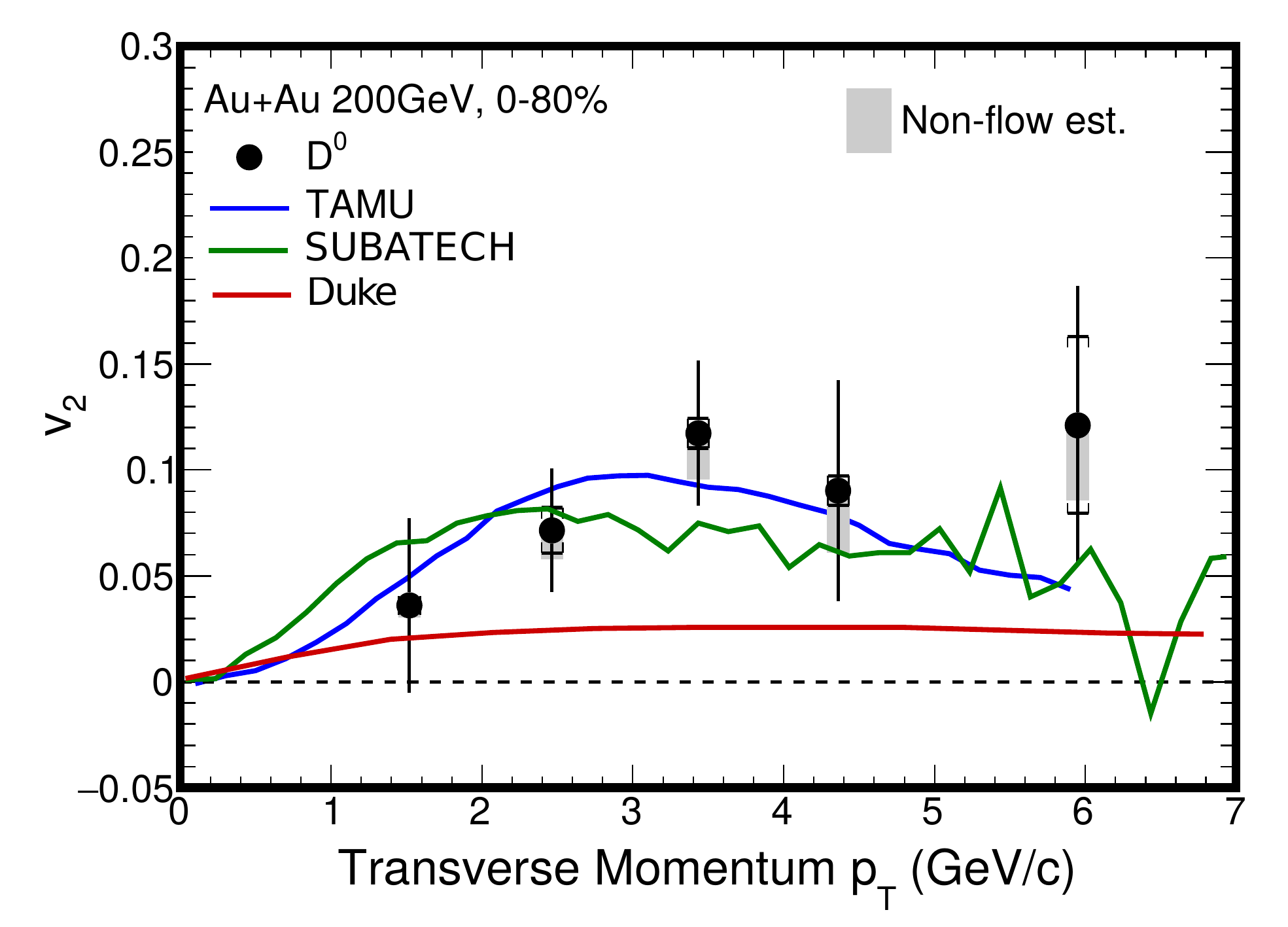}
\end{center}
\end{minipage}
\put(-135, 25) {\footnotesize \color{blue} STAR Preliminary}
\caption{ $D^0$ $v_2$ compared to various models.}
\label{fig:d0v2_2}
\end{figure}

Figures 7 and 8 show the measured $D^0$ \raa for 0-10\% central \AuAu collisions and $v_2$ for 0-80\% \AuAu collisions compared with various model calculations. The Duke model uses a Langevin simulation with an input charm quark diffusion coefficient parameter- (2${\pi}TD_{S}$) fixed to 7, where $D_S$ is the charm quark spacial diffusion coefficient and $T$ is medium temperature. The parameter in the DUKE model is tuned to the LHC $D$-meson \raa data~\cite{theory, duke}. The TAMU calculation uses a non-perturbative approach and the full T-matrix calculation with the internal energy as the potential, which predicts 2${\pi}TD_{S}$ to be $\sim$ 3-11~\cite{theory}. The SUBATECH group uses the $MC@sHQ$ calculation with the latest EPOS3 initial conditions and the resulting 2${\pi}TD_{S}$ $\sim$2-4~\cite{theory}. These three models can describe the measured $D^0$ \raa reasonably well. Meanwhile, the TAMU and SUBATECH calculations can describe the measured $D^0$ $v_2$ as well, while the DUKE calculation with 2${\pi}TD_{S}= 7$ underestimates the $D^0$ $v_2$. To further constrain the medium diffusion coefficient, it will be beneficial to systematically study the effect of each ingredient in different model calculations. 

\begin{figure}[htbp]
\hspace{+1.2cm}
\begin{minipage}[b]{0.80\linewidth}
\begin{center}
\includegraphics[width=\textwidth]{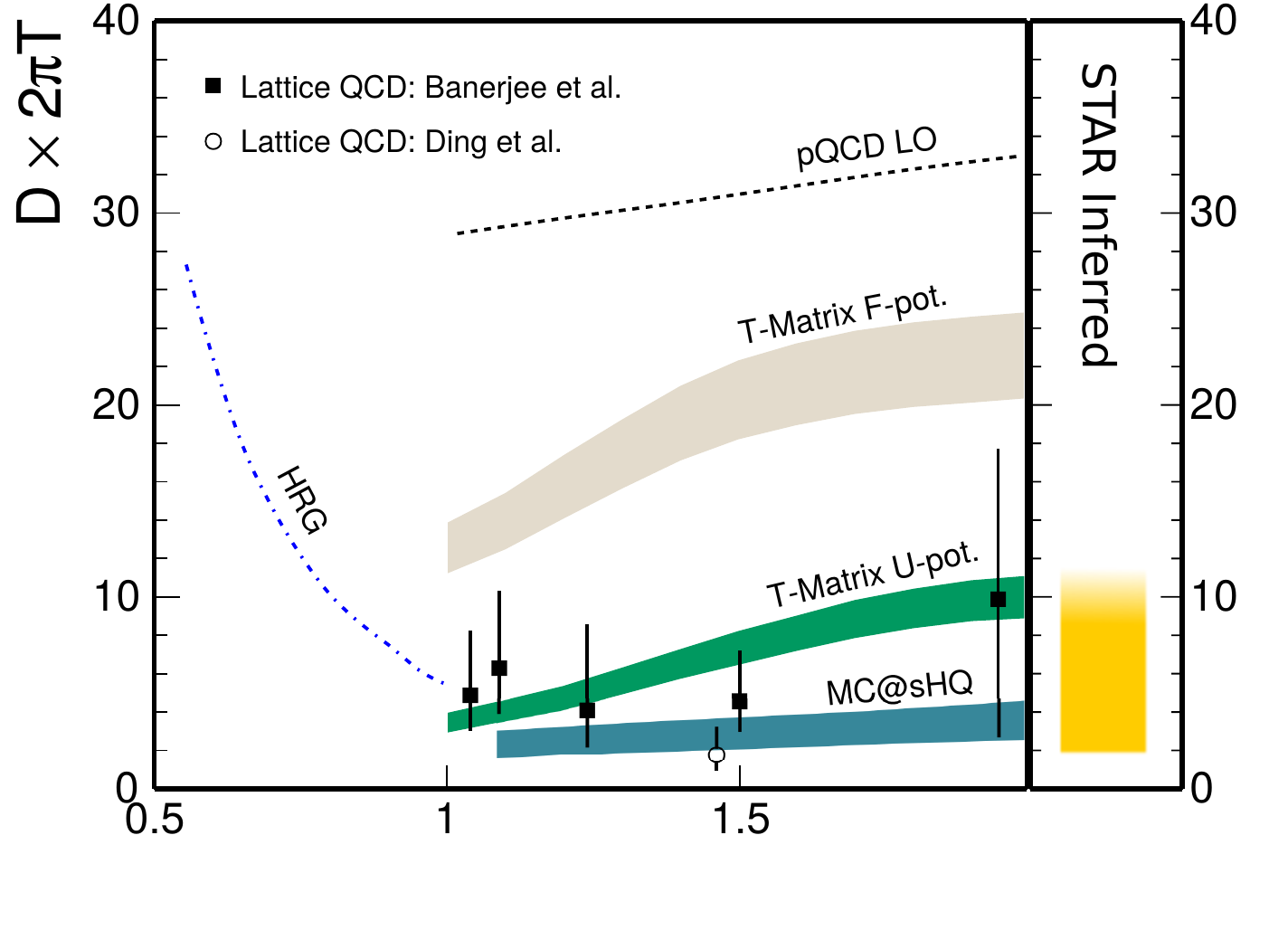}
\end{center}
\end{minipage}
\put(-137, 88) {\footnotesize \color{blue} STAR Preliminary}
\caption{ Charm quark diffusion coefficient from models and the inferred range from STAR measurements.}
\label{fig:diffusioncoefficient}
\end{figure}

Figure 9 shows the extracted diffusion coefficient from different model calculations compared to the yellow band indicating the inferred values from current measurements.

\section{Summary and Outlook}

We report the first measurement of the $D^0$ \raa in the most central 0-10\% centrality bin and $v_2$ in the 0-80\% centrality bin in \AuAu collisions at $\sNN$ = 200 GeV using the recently installed HFT at the STAR experiment. The new \raa results confirm the strong suppression of the $D^0$ yield at high \pT with much improved precision. The measured $D^0$ $v_2$ is non-zero and systematically below the $v_2$ of light hadrons in 0-80\% \AuAu collisions. Theoretical models with charm quark diffusion coefficient 2${\pi}TD_{S}$ $\sim$ 2-12 can reproduce simultaneously the measured $D^0$ \raa and $v_2$ in \AuAu collisions at RHIC. 

A factor of 2-4 improvement in the $D^0$ signal significance is expected from the reprocessed Run 2014 data. With additional 2 billion minimum-bias events taken in Run 2016 with full Al-cables, another factor of 2-3 further improvement is expected. These improvements will allow precise measurements of the centrality dependence of the $D_0$ \raa and $v_2$ in the near future.

\section*{Acknowledgement}
We express great gratitude to RNC group at LBNL and HEP group at USTC for their support. From USTC, the author is supported in part by the NSFC under Grant No.s 11375172, 11375184 and 11675168, and MoST of China under No. 2014CB845400. 




\bibliographystyle{elsarticle-num}
\bibliography{jos}







\end{document}